\documentstyle{edbk}

\input epsf

\newcommand{\dir}{.}
\newcommand{\fig}[3]
{
     \noindent
     \unitlength=1mm
     \begin{picture}(#2,#3)
     \put(0,0){\leavevmode \epsfxsize=#2mm \epsffile{\dir/#1}}
     \end{picture}
   \noindent
}
\normallatexbib
\begin{document}
\articletitle{Order and Disorder Phenomena at Surfaces of Binary Alloys}
\author{F. F. Haas}
\affil{Institut f\"ur Physik, Johannes Gutenberg Universit\"at Mainz, D55099 Mainz}
\author {F. Schmid}
\affil{Max-Planck Institut f\"ur Polymerforschung, Ackermannweg 10, D55021 Mainz}
\email {schmid@mpip-mainz.mpg.de} 
\author{K. Binder}
\affil{Institut f\"ur Physik, Johannes Gutenberg Universit\"at Mainz, D55099 Mainz}
\email {binder@chaplin.physik.uni-mainz.de}

\begin{abstract}
We present recent Monte Carlo results on surfaces of bcc-structured binary 
alloys which undergo an order-disorder phase transformation in the bulk. 
In particular, we discuss surface order and surface induced disorder 
at the bulk transition between the ordered (DO${}_3$) phase and
the disordered (A2) phase. An intricate interplay between different ordering 
and segregation phenomena leads to a complex surface behavior, 
which depends on the orientation of the surface under consideration.
\end{abstract}

The structure and composition of alloys at external surfaces and internal 
interfaces often differs significantly from that in the bulk. 
In most cases, this refers only to very few top layers at the surface, over 
a thickness of order 1 nm. In the vicinity of a bulk phase transition, however,
the thickness of the altered surface region can grow to reach mesoscopic 
dimensions, of order 10-100 nm. If the bulk transition is second order, for
example, the thickness of the surface region is controlled by the bulk
correlation length, which diverges close to the critical point\cite{surf}.
Close to first order bulk transitions, mesoscopic wetting layers may 
form\cite{wetting}.

While these various surface phenomena are fairly well understood in simple
systems, such as surfaces of liquid mixtures against the wall of a 
container, the situation in alloys is complicated due to the interplay 
between the local structure, the order and the composition profiles.
In alloys which undergo an order/disorder transition, for example, 
the surface segregation of one alloy component can induce surface 
order\cite{ich1,dosch2,diehl1} or partial surface order\cite{reichert,ich2}
at surfaces which are less symmetric than the bulk lattice with respect to 
the ordered phase. Even more subtle effects can lead to surface order at
fully symmetric surfaces\cite{mailander,schweika1,schweika2,schweika3}. 
Furthermore, different types of order may be present in such 
alloys\cite{defontaine}, which can interact in a way to affect the
wetting behavior significantly\cite{hauge,gerhard}. 

In this contribution, we discuss a situation where such an interplay of 
segregation and different types of ordering leads to a rather intriguing
surface behavior: Surface induced disorder in a binary (AB) alloy
on a body centered cubic (bcc) lattice close to the first order bulk transition 
between the ordered DO${}_3$ phase and the disordered bulk phase.
Surface induced disorder is a wetting phenomenon, which can be observed when
the bulk is ordered and the surface reduces the degree of ordering 
-- usually due to the reduced number of interacting 
neighbors\cite{lipowsky1,kroll,dosch1}. A disordered layer may then nucleate at 
the surface, which grows logarithmically as the bulk transition is approached. 
According to the theoretical picture, the surface behavior is driven by
the depinning of the interface between the disordered surface layer and the 
ordered bulk. 

In order to study the validity of this picture, we consider a very idealized 
minimal model of a bcc alloy with a DO${}_3$ phase: The alloy is mapped on 
an Ising model on the bcc lattice with negative nearest and next nearest
neighbor interactions. The Hamiltonian of the system then reads
\begin{equation}
{\cal H} =  V \sum_{\langle ij \rangle} S_i S_j
+ \alpha V \sum_{\langle \langle ij \rangle \rangle } S_i S_j
- H \sum_i S_i,
\end{equation}
where Ising variables $S=1$ represent A atoms, $S=-1$ B atoms, the
sum $\langle ij \rangle$ runs over nearest neighbor pairs, 
$\langle \langle ij \rangle \rangle$ over next nearest neighbor pairs, and the 
field $H$ is the appropriate combination of chemical potentials 
$\mu_A$ and $\mu_B$ driving the total concentration $c$ of A in the alloy,
($c= (\langle S \rangle+1)/2$). 

The phases exhibited by this model are shown in Figure 1
In the disordered (A2) phase, the A and B particles are distributed 
evenly among all lattice sites. In the ordered B2 and DO${}_3$ phases, they 
arrange themselves as to form a superlattice on the bcc lattice. The
parameter $\alpha$ was chosen $\alpha=0.457$, such that the highest 
temperature at which a DO${}_3$ phase can still exist is roughly half the 
highest temperature of the B2 phase, like in the experimental case
of FeAl. The resulting phase diagram is shown in Figure 2.

In order to characterize the ordered phases, it is useful to divide the bcc 
lattice into four face centered cubic (fcc) sublattices as indicated on 
Figure 1, and to define the order parameters
\begin{eqnarray}
\psi_1 &=& \big( 
     \langle S \rangle_a + \langle S \rangle_b 
     - \langle S \rangle_c - \langle S \rangle_d )/2 \nonumber \\
\psi_2 &=& \big( \langle S \rangle_a - \langle S \rangle_b 
     + \langle S \rangle_c - \langle S \rangle_d )/2  \\
\psi_3 &=& \big( \langle S \rangle_a - \langle S \rangle_b 
     - \langle S \rangle_c + \langle S \rangle_d )/2, \nonumber
\end{eqnarray}
where $\langle S \rangle_{\alpha}$ is the average spin on the sublattice
$\alpha$. In the disordered phase, all sublattice compositions are equal
and all order parameters vanish as a consequence. The B2 phase is
characterized by $\psi_1 \ne 0$ and the DO${}_3$ phase by
$\psi_1 \ne 0$ and $\psi_2 = \pm \psi_3 \ne 0$. The two dimensional
vector $(\psi_2,\psi_3)$ is thus an

\begin{minipage}[b]{6.cm}
\fig{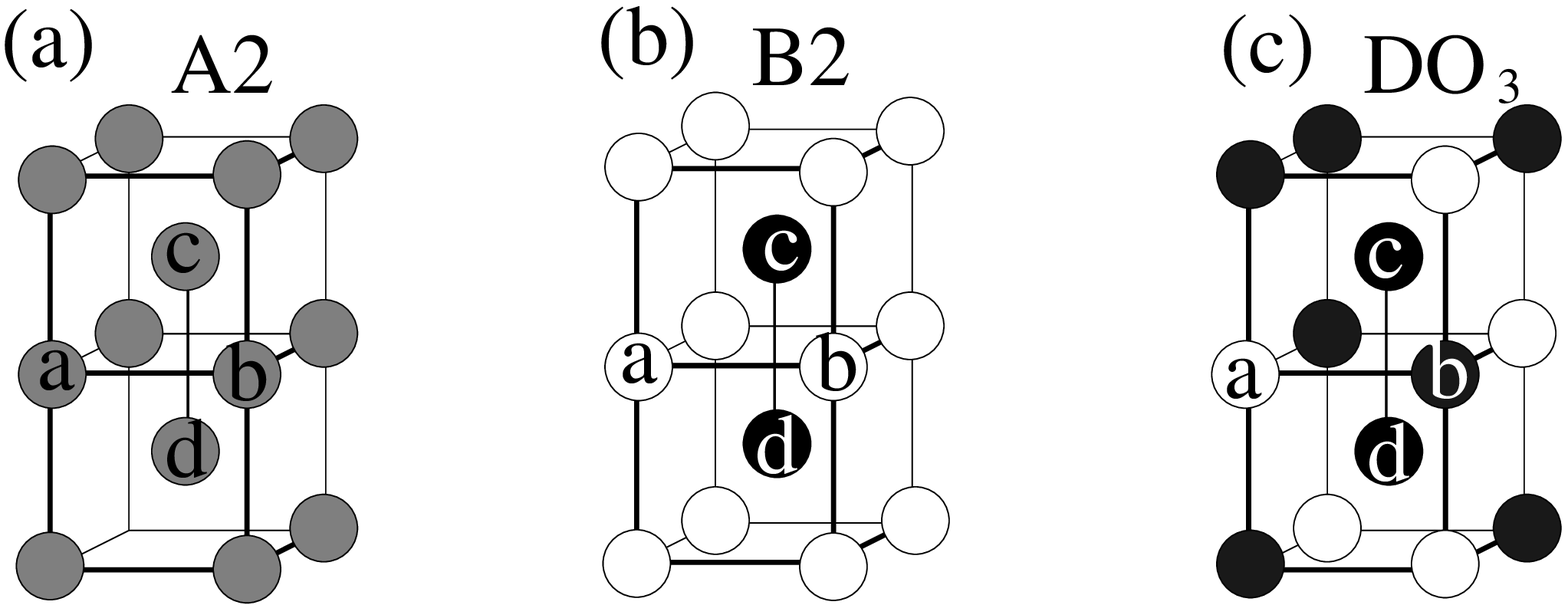}{55}{30}
\vspace{1.cm}
\baselineskip=10pt
Figure 1.\\
{\small 
Ordered phases on the bcc lattice: (a) disordered A2 structure, 
(b) ordered B2 and (c) DO${}_3$ structure.
Also shown is assignment of sublattices $a,b,c$ and $d$.
}
\end{minipage}
\hfill
\begin{minipage}[b]{6.cm}
\fig{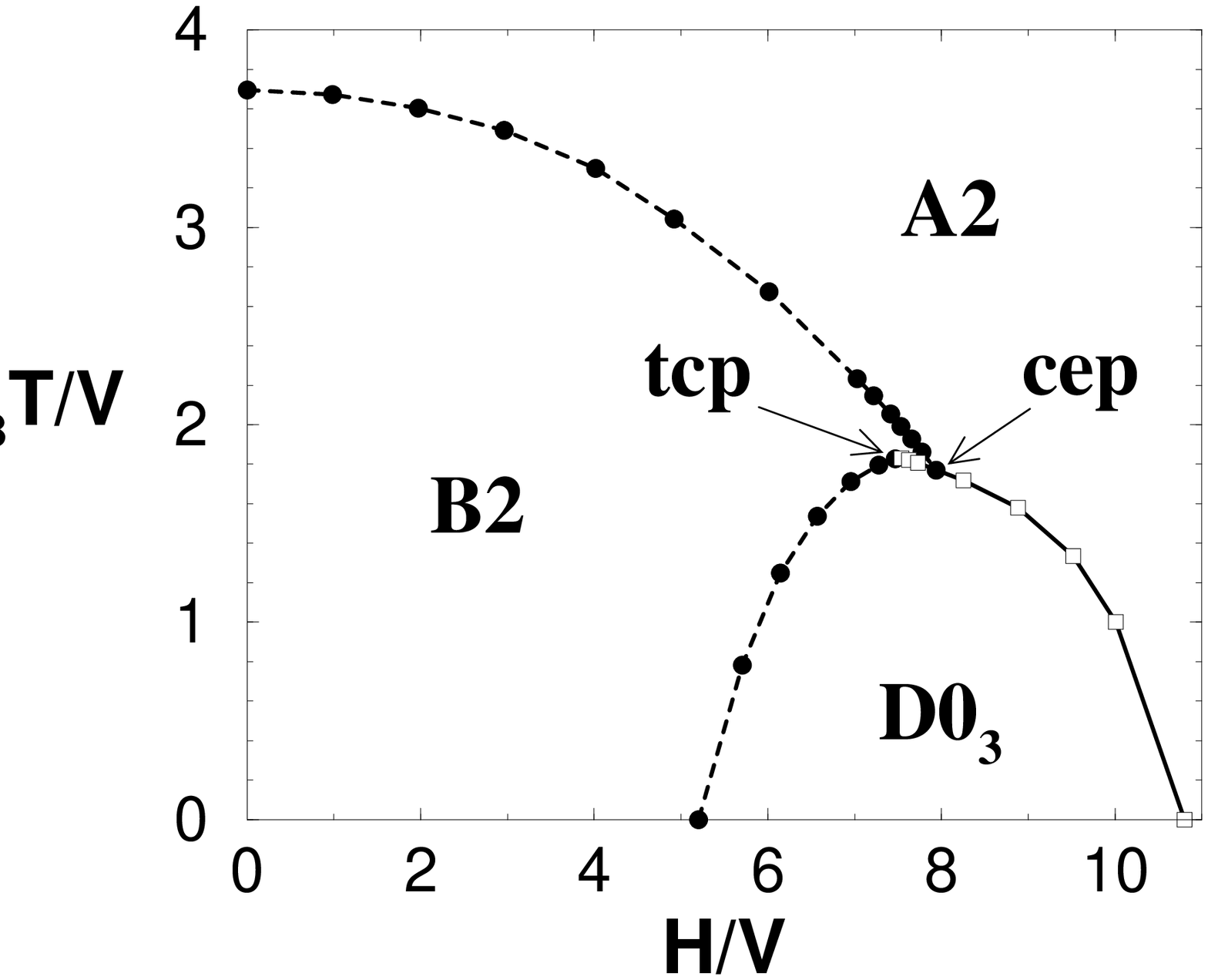}{55}{60}
\baselineskip=10pt
Figure 2.\\
{\small 
Phase diagram in the $T-H$ plane. First order transitions are solid lines,
second order transitions dashed lines. Arrows indicate positions of a
critical end point (cep) and a tricritical point (tcp).
}
\end{minipage}\\
\bigskip
\noindent

\noindent
order parameter for DO${}_3$
ordering, and the latter can be characterized conveniently in terms of
its absolute value
\begin{equation}
\psi_{23} = \sqrt{(\psi_2{}^2 + \psi_3{}^2)/2}.
\end{equation}

We have studied free (110) and (100) surfaces of this model at the temperature 
$T = 1\: k_B T/V$ in the DO${}_3$ phase close to the transition to the
disordered phase. To this end, we have first located the transition point
very accurately by thermodynamic integration\cite{KB2}, $H_0/V=10.00771[1]$.
We have then performed extensive Monte Carlo simulations of slabs each
100-200 layers thick, with free boundary conditions at the two 
confining (110) or (100) planes, and periodic boundary conditions in the 
remaining directions. 

In all of our simulations, the average value of the Ising variable in the
top layer was one, {\em i.e.}, the top layer was completely filled with A 
atoms. Having stated this, we shall disregard this layer in the following and
discuss the structure starting from the next layer underneath the surface. 
The layer order parameters $\psi_i(n)$ and the layer compositions $c(n)$ 
can be determined in a straightforward manner for (110) layers, since they
contain sites from all sublattices. 
In the case of the (100) layers, it is useful to define $c$ and the $\psi_i$
based on the sublattice occupancies on two subsequent layers.

\bigskip

Fig. 3 shows the calculated profiles for two choices of $H$ close to the 
transition. One clearly observes the formation of a disordered film at the 
surface, which increases in thickness as the transition point is approached. 
The film is characterized by low order parameters $\psi_1$ and $\psi_{23}$,
and by a slightly increased concentration $c$ of A sites. The structure very
close to the surface depends on its orientation: The composition profiles
display some characteristic oscillations at a (110) surface, and grow
monotonously at a (100) surface. The order $\psi_{23}$ drops to zero. 
The order $\psi_1$ drops to zero at the (110) surface, and at
the (100) surface, it changes sign and increases again in the outmost two
layers. The latter is precisely an example of
the segregation induced ordering mentioned earlier\cite{ich1,dosch2,diehl1}.

\bigskip
\bigskip

\noindent
\begin{minipage}[t]{13cm}
(110) surface \hfill (100) surface \\
\fig{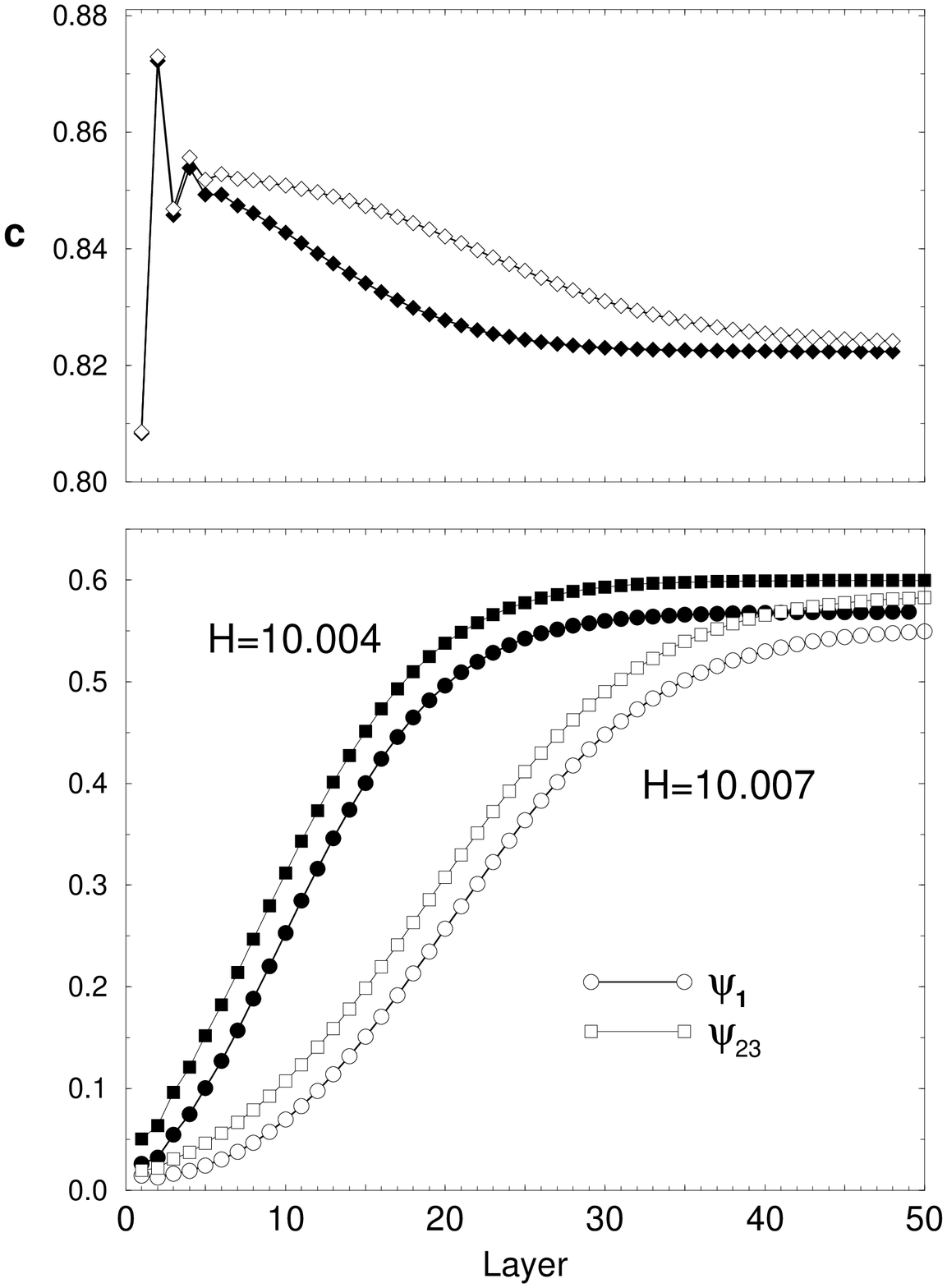}{55}{85} \hfill 
\fig{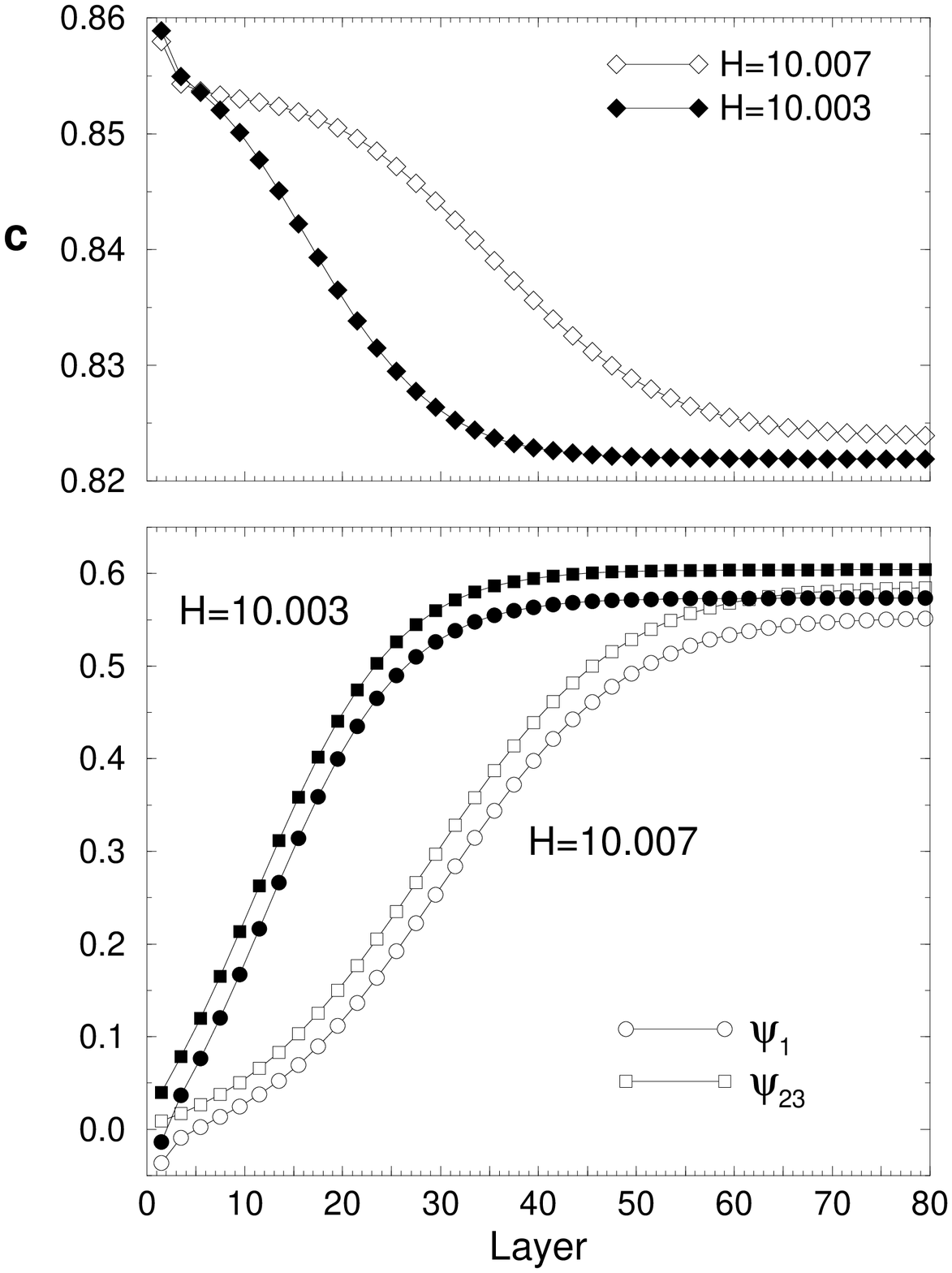}{55}{85} \\
\baselineskip=10pt
Figure 3.\\
{\small 
Profiles of the total concentration (top) and order parameters $\psi_1$
(bottom, circles) and $\psi_{23}$ (bottom, squares) at the
(110) surface (left) and at the (100) surface (right) for different
fields $H$ in units of $V$ as indicated.
}
\end{minipage}\\
\bigskip

We can thus distinguish between two interesting regions in these profiles:
The near-surface region, where the properties of the profiles still
reflect the peculiarities of the surface,
and the interfacial region, where the profiles are determined from the
properties of the interface separating the disordered surface film 
from the ordered bulk.

The structure of the profiles in the interfacial region is basically
determined by the fluctuations of the interface, which are characterized
by a transverse correlation length $\xi_{\parallel}$. The latter is in turn 
driven by the thickness of the film and the interfacial tension $\sigma$ 
or, more precisely, by a rescaled dimensionless interfacial tension
\begin{equation}
1/\omega =  4 \pi \xi_b{}^2 \sigma/k_B T,
\end{equation}
with the bulk correlation length $\xi_b$. The renormalization group theory
of critical wetting\cite{wetting}, which should apply here\cite{kroll}, 
predicts that the transverse correlation length diverges according 
to a power law
\begin{equation}
\xi_{\parallel} \propto \frac{1}{\sqrt{\omega}} (H_0 - H)^{-\nu_{\parallel}}
\end{equation}
as $H_0$ is approached, with the exponent $\nu_{\parallel} = 1/2$.
With this knowledge, one can calculate the effective width
of the order parameter profiles, 
$\xi_{\perp} \propto \sqrt{- \omega \ln(H_0-H)}$, the profiles of
layer susceptibilities etc. We have examined these carefully at
the (110) interface and the (100) interface, both for the order
parameters $\psi_1$ and $\psi_{23}$, and we could fit everything nicely
into the theoretical picture. 
Our further discussion here shall focus on the near-surface region.

Assuming that the order in the near-surface region is still determined by the 
fluctuations of the interface, the theory of critical wetting 
predicts a power law behavior
\begin{equation}
\label{b1}
\psi_{\alpha,1} \propto (H_0 - H)^{\beta_1} 
\end{equation}
for value $\psi){\alpha,1}$ of the order parameter $\psi_{\alpha}$ directly 
at the surface, regardless of the structure of the surface. Figure 4 shows 
that, indeed, the surface order parameter $\psi_{23,1}$ decays according to 
a power law at both the (110) and the (100) surface, with the same exponent
and the exponent is in both cases identical within the error,
$\beta_1 = 0.618$. Furthermore, we notice strong finite size effects 
close to the $H_0$. Since these are asymptotically driven by 
the ratio $(L/\xi_{\parallel})$, they can be exploited to determine
the behavior of $\xi_{\parallel}$ as the phase transition is approached, 
{\em i.e.}, the exponent $\nu_{\parallel}$. The finite size scaling analysis 
yields $\nu_{\parallel}=1/2$\cite{frank}, 
in agreement with the theory.

\bigskip
\bigskip

\noindent
\begin{minipage}[t]{13.cm}
(110) surface \hfill (100) surface \\
\fig{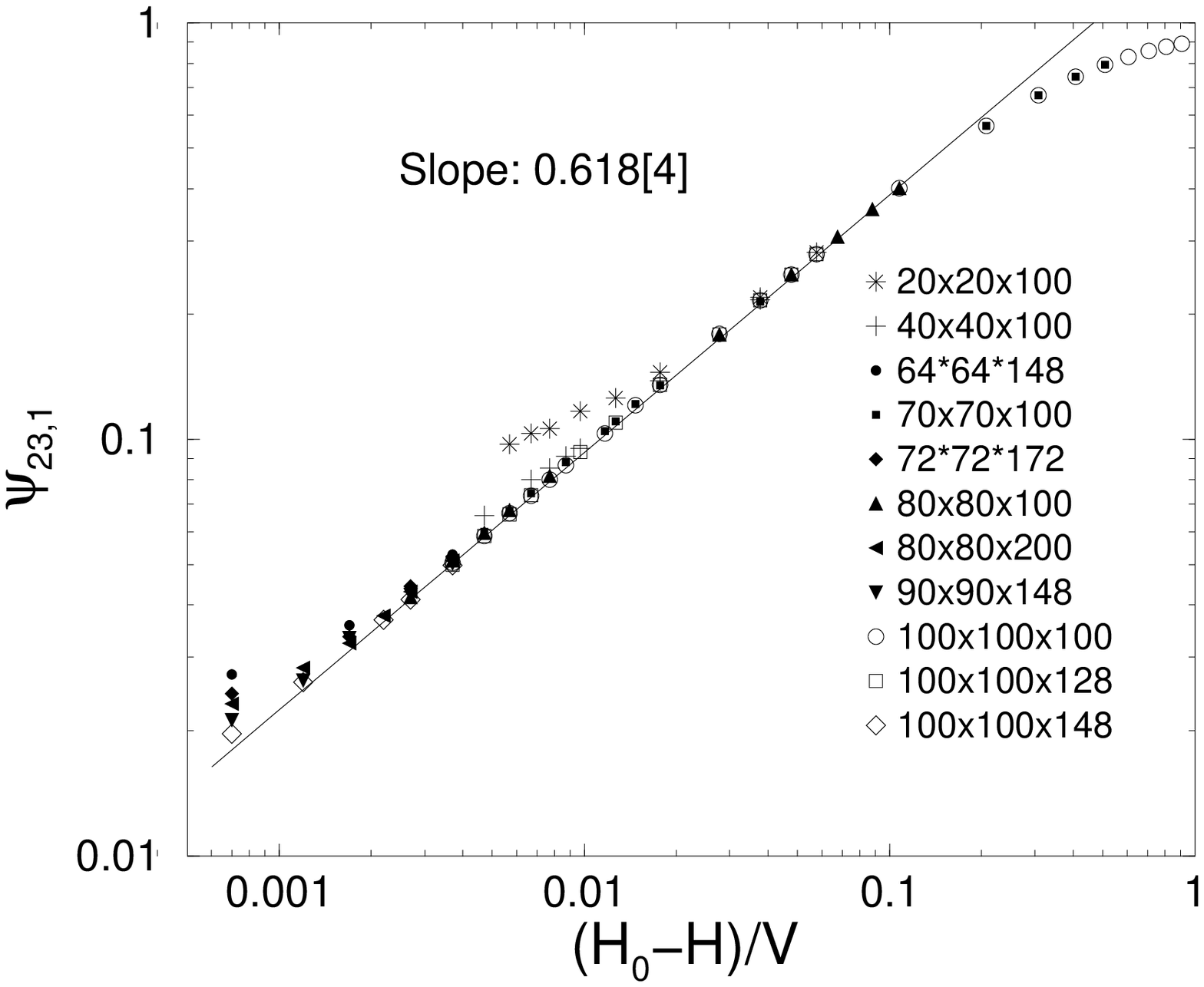}{55}{50} \hfill \fig{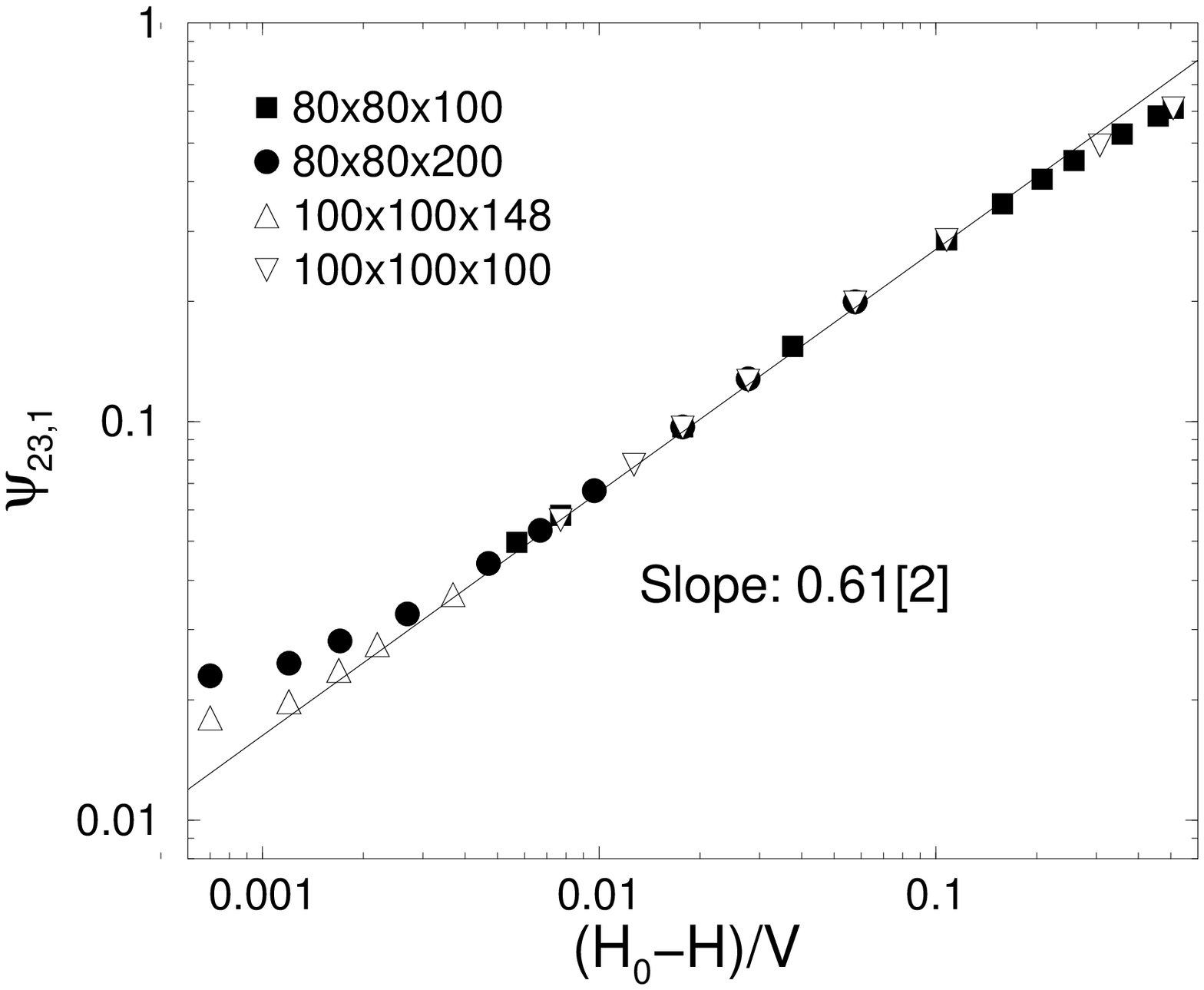}{55}{50}\\
\baselineskip=10pt
Figure 4.\\
{\small 
Order parameter $\psi_{23}$ at the surface at the (110) surface (left)
and the (100) surface (right) vs. $(H_0-H)/V$ for different
system sizes $L\times L \times D$ as indicated. Solid line shows power
law with exponent $\beta_1 = 0.618$.
}
\end{minipage}\\
\bigskip

The profiles of $\psi_{23}$ thus seem entirely determined by the depinning
of the interface, in agreement with the standard theory of critical wetting.
The situation is however different when one looks at the other order parameter,
$\psi_1$. This is not particularly surprising in the case of the (100) surface.
We have already noted that this surface breaks the symmetry with respect
to the $\psi_1$ ordering, hence the segregation of A particles to the
top layer induces additional $\psi_1$ order at the surface (Figure 5). 
The (110) surface, on the other hand, is not symmetry breaking. 
The order $\psi_1$ decays at the surface, yet with an exponent
$\beta_1 = 0.801$ which differs from that observed for $\psi_{23}$ (Figure 6).
Even more unexpected, the finite size effects cannot be analyzed consistently 
with the assumption that the transverse correlation length diverges with the 
exponent $\nu_{\parallel}=1/2$, but rather suggest 
$\nu_{\parallel} = 0.7 \pm 0.05$. The order parameter fluctuations of 
$\psi_1$ at the surface seem to be driven 
by a length scale which diverges at $H_0$ with an exponent different from
that given by the capillary wave fluctuations of the depinning interface.

\bigskip
\bigskip

\noindent
\begin{minipage}[t]{6.cm}
\fig{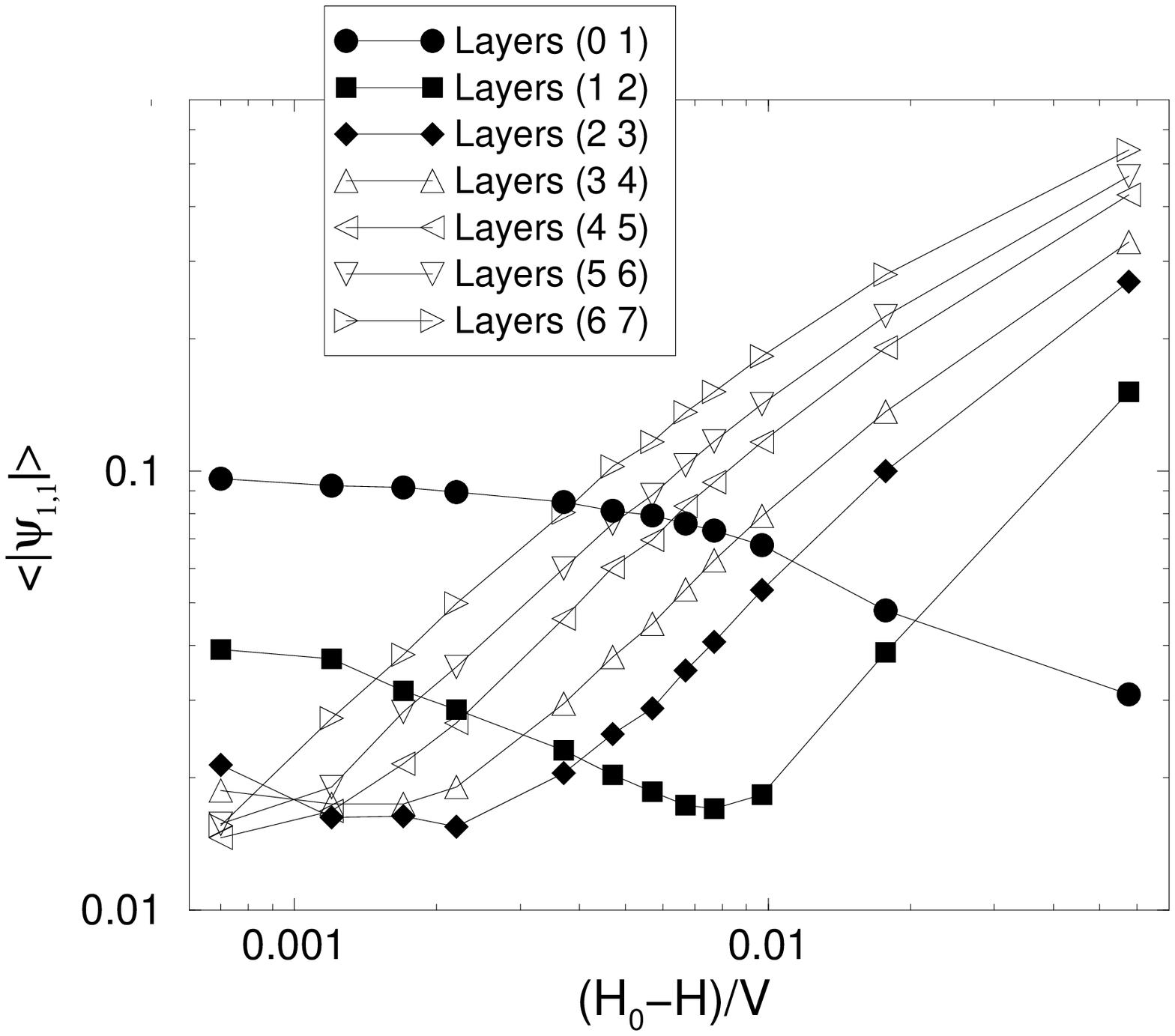}{55}{50}
\baselineskip=10pt
Figure 5.\\
{\small 
Order parameter $\psi_{1}$ in the first layers underneath
the (100) surface vs. $(H_0-H)/V$ .
}
\end{minipage}
\hfill
\begin{minipage}[t]{6.cm}
\fig{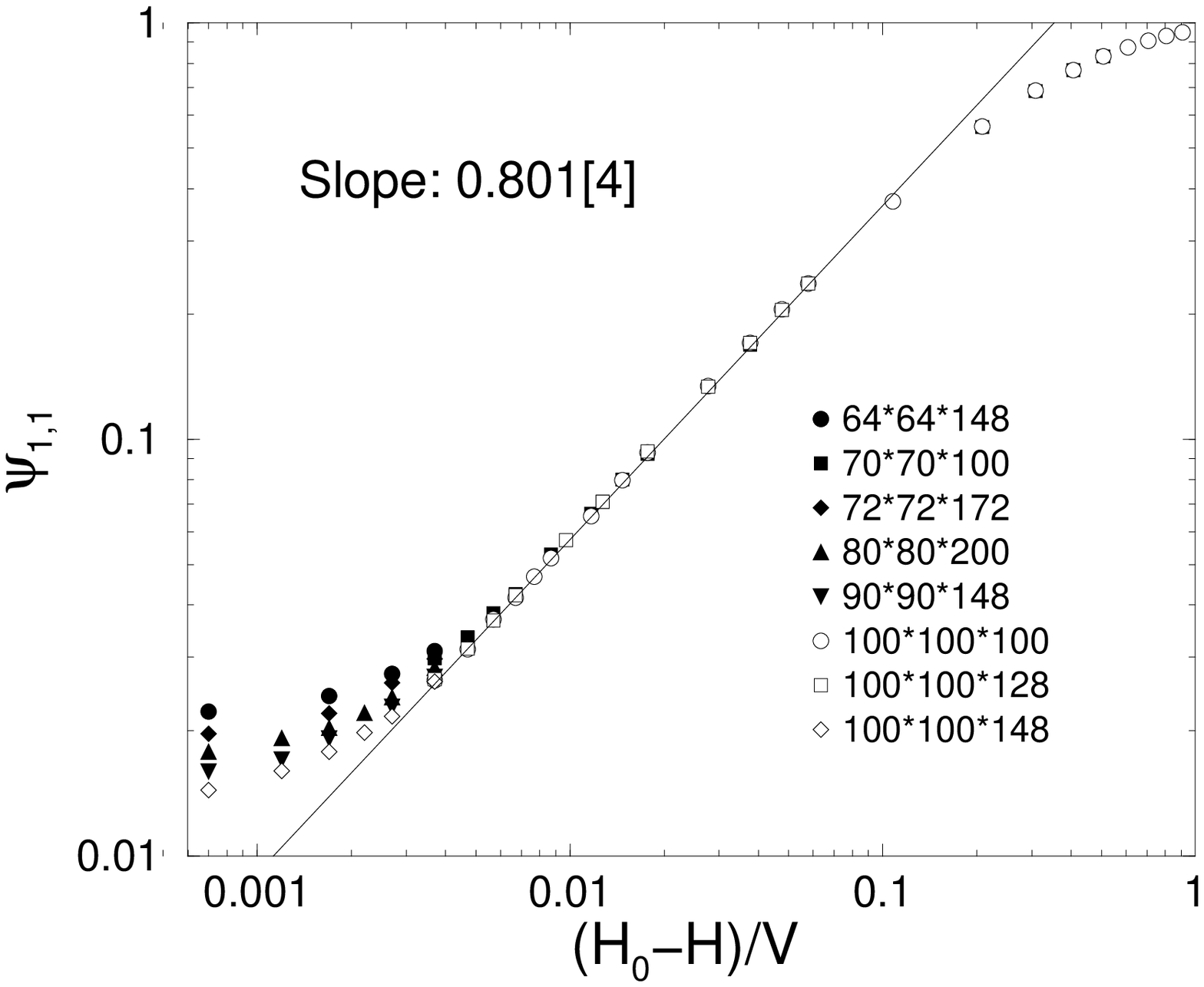}{55}{50}
\baselineskip=10pt
Figure 6.\\
{\small 
Order parameter $\psi_{1}$ at the (110) surface 
vs. $(H_0-H)/V$ for different
system sizes $L\times L \times D$ as indicated. 
Solid line shows power
law with exponent $\beta_1 = 0.801$.
}
\end{minipage}

\bigskip

To summarize, we have seen that the phenomenology of surface induced disorder 
in a relatively simple bcc alloy with just two coupled types of ordering 
is much more complex than predicted by the standard theory of surface induced
disorder and critical wetting.
Looking at our profiles, we were able to distinguish between two regions, 
the near-surface region and the interfacial regions. 
In the situations studied in our simulations, it seemed that these regions 
could be well separated from each other. 
Some rapid variations of the profiles in the near-surface region,
are followed by smooth changes in the interfacial region. 
The local surface structure affects the total composition profile relatively
strongly, and has practically no influence on the profile of the DO${}_3$
order, $\psi_{23}$. In the case of the B2 order, $\psi_1$, the situation
is more intriguing. The symmetry breaking (100) surface induces local order 
in the near-surface region which apparently does not couple to the interface.
At the non-symmetry breaking (110) surface, $\psi_1$ was found to
exhibit qualitatively new and unexpected power law behavior as the
wetting transition is approached. 

The last observation clearly requires further exploration in the future.
The picture will be even more complex in situations
where the near-surface profiles and the interfacial profiles cannot be
separated any more. We expect that this could be the case, {\em e.g.},
at (111) surfaces, which break the symmetry with respect to both
B2 and DO${}_3$ ordering.

\bigskip

\noindent
\small
\baselineskip=11pt
F.F. Haas was supported by the Graduiertenf\"orderung of the Land 
Rheinland-Pfalz.

\end{document}